\documentclass[twocolumn]{article}
\usepackage[margin=16mm]{geometry}
\usepackage{sectsty}
\sectionfont{\normalsize}
\subsectionfont{\normalsize}
\usepackage{authblk}
\usepackage[numbers,sort&compress]{natbib}
\usepackage{siunitx}
\usepackage{amsmath}
\usepackage{amsfonts}
\usepackage{amssymb}
\usepackage{graphicx}
\usepackage[font=small]{caption}
\usepackage{hyperref}
\hypersetup{
    colorlinks=true,
    linkcolor=blue,
    filecolor=magenta,      
    urlcolor=cyan,
    }

\newcommand{\Pe}{{\rm Pe}}
\newcommand{\Dreff}{D_r^{\rm eff}}

\providecommand{\keywords}[1]
{
  \small	
  \textbf{\textit{Keywords---}} #1
}

\title{
Flow-history-dependent orientational relaxation in dilute polydisperse colloidal rod suspensions
}

\author[1,2]{Yuto Yokoyama}

\author[1,3]{Vincenzo Calabrese}

\author[1]{Fabian Hillebrand}

\author[1,4]{Henry J. London}

\author[1]{Simon J. Haward}

\author[1]{Amy Q. Shen}

\affil[1]{
    Micro/Bio/Nanofluidics Unit, Okinawa Institute of Science and Technology Graduate University, Japan}

\affil[2]{
    Physical Science and Engineering Division, King Abdullah University of Science and Technology, Kingdom of Saudi Arabia}

\affil[3]{
    POLYMAT, Rheology and Advanced Manufacturing group, University of the Basque Country UPV/EHU, Spain}

\affil[4]{
    Laboratory of Systems and Synthetic Microbiology, Department of Engineering, University of Cambridge, United Kingdom}

\begin{document}

\date{}

\twocolumn[

\maketitle

\begin{abstract}
Orientation and relaxation dynamics of rod-like colloids under flow govern the optical and mechanical properties of many emerging soft materials.
    In polydisperse suspensions, particles of different lengths exhibit distinct rotational diffusion timescales, yet how this polydispersity influences relaxation following flow cessation remains unclear.
    In particular, it is not well understood how the pre-shear rate determines the subsequent orientation relaxation dynamics.
    To address this question, we performed simple shear on dilute cellulose nanocrystal (CNC) suspensions in a narrow-gap Taylor-Couette cell and measured birefringence relaxation after flow cessation using high-speed polarization imaging.
    To interpret the experiments, we formulated a polydisperse Fokker-Planck model parameterized by the measured length distribution.
    As a result, the average orientation relaxation time systematically decreases with increasing pre-shear rate.
    Moreover, when organized by the P\'{e}clet number based on the rotational diffusion coefficient of the weighted average rod length, the data agree well with the theory over a wide range of shear rates.
    This trend arises because the rod sub-population contributing most strongly to the orientation shifts from longer rods to shorter rods as the pre-shear rate increases, showing that the flow history governs the orientation relaxation dynamics.
    In polydisperse systems, the orientation relaxation time is no longer a material-specific constant but is determined by both the flow conditions and the polydispersity.
    This study provides a quantitative framework for understanding orientation dynamics in polydisperse rod suspensions and for interpreting rheo-optical measurements.
\end{abstract}

\vspace{3mm}

\keywords{
Rod orientation, 
Polydispersity, 
Relaxation time, 
Cellulose nanocrystals, 
Flow birefringence
}

\vspace{8mm}
]

\section{Introduction}


Rod-like colloidal particles play a central role in advanced soft materials because their orientational dynamics under flow directly control macroscopic optical, mechanical, and rheological properties \citep{butler2018}.
In this work, we focus on dilute suspensions of rigid Brownian rods and use cellulose nanocrystals (CNCs), which are highly crystalline, rod-shaped nanoparticles derived from wood and plant biomass, as a well-characterized model system \citep{habibi2010,abbasimoud2023}.
CNCs form lyotropic liquid-crystalline phases and serve as fundamental components for optical films, food materials, and 3D-printing inks \citep{gauss2021,lagerwall2014,mu2019}.
Because these applications rely sensitively on flow-controlled alignment, understanding how rods orient under imposed deformation is essential, particularly in polydisperse suspensions and under unsteady flow conditions.


The orientational dynamics of rod-like particles in flow are commonly described by theories that couple hydrodynamics to rotational motion \citep{doi1988}. 
At the single-particle level, Jeffery's equation predicts the tumbling and alignment of an axisymmetric rod \citep{jeffery1922,leal1972}. 
At the suspension level, the evolution of the probability density function for particle orientation is governed by a Fokker-Planck equation balancing rotational advection and diffusion \citep{doi1988,talbot2024}.
From this probability density function, orientational order can be quantified and directly compared with experimental optical measurements such as flow-induced birefringence \citep{fuller1995, salipante2025}.

Flow-induced birefringence provides a direct optical probe of the orientational dynamics of rod-like particles \citep{calabrese2023,kadar2021, pritchard2021}. 
A suspension at rest is optically isotropic with randomly oriented rods, whereas an imposed flow aligns the rods toward a preferred direction and thereby generates birefringence \citep{fuller1995}. 
The birefringence magnitude $\Delta n$ reports the degree of orientational order, while the orientation angle $\phi$ reports the principal alignment direction of the rods.
As a result, flow birefringence has long been used to monitor rod alignment in shear and extensional flows across a wide range of rod-like systems, including viruses, rare-earth-doped nanorods, and CNCs \citep{calabrese2024,nakamine2024a,kim2017}.

Transient birefringence in unsteady flows encodes the intrinsic relaxation dynamics of the orientational microstructure, which are governed by rotational diffusion.
A rod reorients on a finite characteristic timescale set by the inverse rotational diffusion coefficient, $D_r^{-1}$ \citep{fuller1995,doi1988}.
This is evident in flow-cessation experiments, where the birefringence decays after an abrupt stoppage of the imposed flow as the orientational microstructure relaxes by rotational diffusion \citep{rosenblatt1985, bellini1989, brouzet2018, rosen2020}.
A recent study has quantified transient birefringence in cross-slot microfluidic devices under sinusoidally modulated planar extensional flow using colloidal rod suspensions \citep{recktenwald2026}.
It revealed that increasing the modulation amplitude and frequency causes the birefringence to deviate from the imposed sinusoidal strain-rate waveform, transitioning to a nonlinear, hysteretic response.
These transient birefringence dynamics provide a sensitive probe of how a given rod population aligns and loses alignment in response to changes in flow.


This picture becomes particularly nontrivial in polydisperse suspensions.
For rod-like colloidal particles such as CNC suspensions, the broad length distribution implies a wide spectrum of rotational diffusion coefficients $D_r(l)\propto l^{-3}$, where $l$ is the rod length \citep{fuller1995,doi1988}. 
This inherent polydispersity complicates the quantitative link between flow conditions, orientational order, and macroscopic response \citep{lang2020,recktenwald2026}.
In polydisperse suspensions, relaxation after flow cessation is therefore inherently multi-modal.
As documented in previous studies \citep{matsumoto1972,brouzet2019,frattini1986}, the measured birefringence reflects contributions from rods of different lengths, and thus a spectrum of relaxation times.
In contrast, a dilute monodisperse suspension is expected to exhibit a single relaxation time $\tau_b = 1/6D_r$, and the birefringence decays exponentially after flow cessation \citep{fuller1995,doi1988}.
These observations have motivated attempts to reconstruct rod length distributions directly from relaxation dynamics, both with and without accounting for particle interactions \citep{arenas-guerrero2018,rogers2005}.
However, rod alignment for different rod lengths depends strongly on the applied shear rate: at low shear rates, primarily the longer rods contribute, while only at higher rates can shorter rods also align. 
Fokker-Planck-type theories attribute this behavior to a P\'{e}clet-number-dependent weighting of different length classes based on the length distribution, where $\Pe=\dot\gamma/D_r$ compares hydrodynamic drift to rotational diffusion of a rod.
As a result, the averaged relaxation time after flow cessation depends not only on the intrinsic length distribution but also on the degree of flow-induced alignment prior to flow cessation.
This suggests that the apparent relaxation time is not an intrinsic material property, but instead emerges from the flow-dependent weighting of rod populations with different rotational diffusivities.

Previous studies have attempted to link rod size and relaxation dynamics from birefringence transients in nonuniform flows, including microchannel configurations where the local strain rate varies along streamlines \citep{brouzet2018,brouzet2019}.
However, in these configurations, the residence time at a given strain rate may be insufficient to reach steady alignment for all length classes, so the extracted ``apparent'' length distribution may reflect partially developed microstructure rather than a fully flow-selected steady state.
As a result, it remains difficult to isolate how steady pre-alignment at a prescribed shear rate selects aligned subpopulations and sets the subsequent relaxation time.


The central question addressed in this work is therefore: how does the shear rate applied prior to flow cessation determine the averaged relaxation time of rod alignment in a dilute polydisperse suspension? 
To address this question, we employ a Taylor-Couette flow cell to bring the suspension to a steady aligned state at each imposed shear rate before cessation. 
This allows us to isolate how flow history selects aligned subpopulations from a known length distribution and thereby governs post-cessation relaxation.
We address this question experimentally using CNC suspensions subjected to steady simple shear and high-speed polarization imaging during flow cessation, and theoretically using a polydisperse Fokker-Planck model parameterized by the measured CNC length distribution.


Because flow birefringence is also often discussed in terms of stress via the stress-optic law (SOL), we briefly note its limitations in the present context.
The stress-optic law states that birefringence is proportional to stress (or strain rate for Newtonian fluids) \citep{noto2020,philippoff1961,lane2023,kim2017}.
In unsteady flows, however, the finite orientational relaxation time can introduce hysteresis, limiting the direct applicability of SOL when the microstructure is out of equilibrium.
For example, \citet{noto2025} showed clear departures from SOL under unsteady conditions in concentrated polymer solutions, highlighting that transient birefringence must be interpreted with care.
Here we focus on using birefringence primarily as a probe of orientational dynamics in dilute rod suspensions.


In the following, we first formulate the orientational Fokker-Planck model for polydisperse rod suspensions (Section~\ref{sec:theory}), then describe the CNC suspensions and Taylor-Couette setup used to probe flow-induced birefringence (Section~\ref{sec:experiment}), and finally compare the theoretical predictions with steady-state and relaxation measurements (Section~\ref{sec:results}).

\section{Theory}
\label{sec:theory}

In this section, we formulate a Fokker-Planck description of the orientational dynamics of rigid Brownian rods under simple shear.
The framework explicitly incorporates rod length polydispersity and provides the basis for analyzing both steady-state alignment and relaxation after flow cessation.

\subsection{Angular velocity of a rod}

\begin{figure}
    \centering
    \includegraphics[width=1\columnwidth]{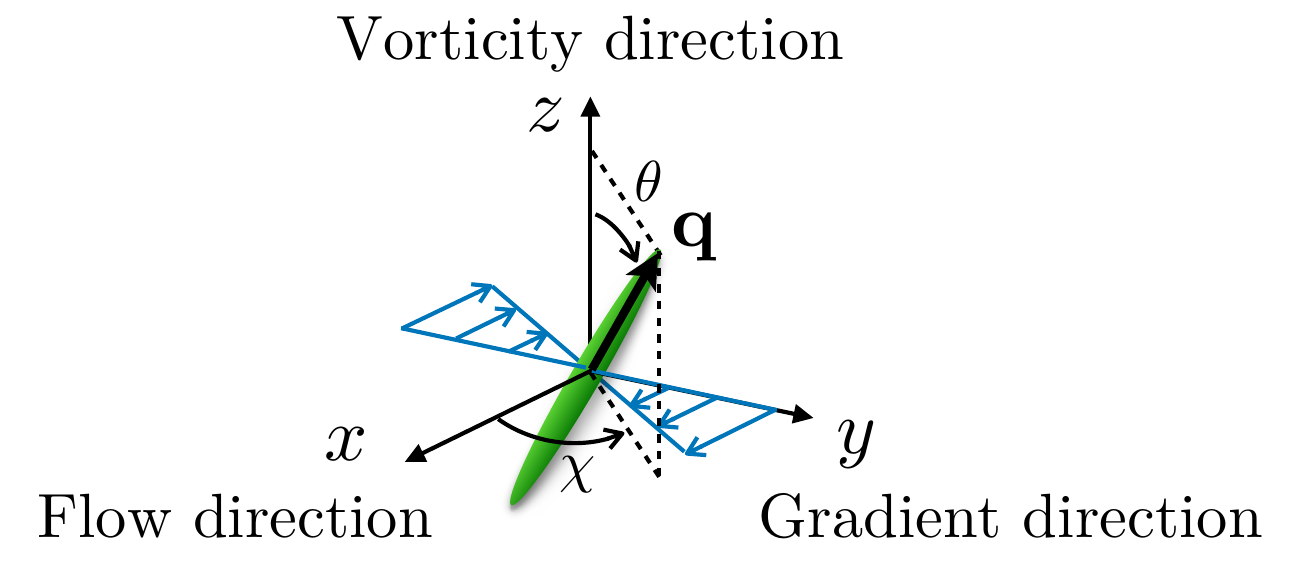}
    \caption{
    Schematic of the coordinate system used to describe rod orientation. 
    The unit vector $\mathbf{q}(\chi,\theta)$ specifies the rod axis, where $\chi$ is the azimuthal angle in the $xy$-plane and $\theta$ is the polar angle measured from the $+z$-axis. 
    A simple shear flow is applied in the $xy$-plane, with flow in the $x$-direction and velocity gradient in the $y$-direction.
    }
    \label{fig:coordinate}
\end{figure}

We begin by defining the coordinate system and the mathematical representation of rod orientation used in this study.
We describe rod orientation on the unit sphere in a Cartesian laboratory frame $(x,y,z)$.
A rigid, axisymmetric rod is described by the unit vector:
\begin{equation}
    \mathbf{q}(\chi,\theta)
    =
    \begin{bmatrix}
    q_x\\
    q_y\\
    q_z
    \end{bmatrix}
    =
    \begin{bmatrix}
    \sin\theta\cos\chi\\
    \sin\theta\sin\chi\\
    \cos\theta
    \end{bmatrix},
\end{equation}
where $\chi$ is the azimuthal angle in the $xy$-plane and $\theta$ is the polar angle measured from the $+z$-axis as shown in Fig.~\ref{fig:coordinate}.
We denote the angular coordinates collectively as $\mathbf{\Omega}=(\chi,\theta)$.

The time evolution of the rod orientation is governed by Jeffery's equation, which balances rotational advection by the flow and the geometric constraint of axisymmetric particles \citep{talbot2024,leal1972}:
\begin{equation} \label{eq:Jeffery}
    \dot{\mathbf{q}} = \mathbf{W} \cdot \mathbf{q} 
    + \beta \left[ \mathbf{E} \cdot \mathbf{q} 
    - (\mathbf{q} \cdot \mathbf{E} \cdot \mathbf{q}) \mathbf{q} \right],
\end{equation}
where $\beta = (r_p^2 -1)/ (r_p^2 + 1)$ is the Bretherton parameter \citep{bretherton1962}, with $r_p=l/d$, where $l$ and $d$ are the length and diameter of the rod particles, respectively.
The Bretherton parameter $\beta \rightarrow 1$ for infinitely slender rods ($r_p \to \infty$), $\beta = 0$ for spheres ($r_p = 1$), and $\beta \rightarrow-1$ for infinitely thin disks ($r_p \to 0$).
$\mathbf{W} = (\nabla \mathbf{u} - \nabla \mathbf{u}^\intercal)/2$ and $\mathbf{E} = (\nabla \mathbf{u} + \nabla \mathbf{u}^\intercal)/2$ denote the vorticity and strain-rate tensors associated with the velocity field $\mathbf{u}=(u_x,u_y,u_z)$.

For a simple shear flow with shear rate $\dot \gamma$ in the $xy$-plane, defined by $u_x=\dot \gamma y$ and $u_y=u_z=0$, $\mathbf{W}$ and $\mathbf{E}$ take the forms:
\begin{equation}
    \mathbf{E} =
    \begin{bmatrix}
    E_{xx} & E_{xy} & E_{xz} \\
    E_{xy} & E_{yy} & E_{yz} \\
    E_{xz} & E_{yz} & E_{zz}
    \end{bmatrix}
    =
    \frac{1}{2}
    \begin{bmatrix}
    0 &  \dot \gamma & 0 \\
     \dot \gamma & 0 & 0 \\ \label{eq: E}
    0 & 0 & 0
    \end{bmatrix},
\end{equation}
\begin{equation}
    \mathbf{W}
    =
    \begin{bmatrix}
    0 & W_{xy} & W_{xz} \\
    -W_{xy} & 0 & W_{yz} \\
    -W_{xz} & -W_{yz} & 0
    \end{bmatrix}
    =
    \frac{1}{2}
    \begin{bmatrix}
    0 & \dot \gamma & 0 \\
    - \dot \gamma & 0 & 0 \\
    0 & 0 & 0
    \end{bmatrix}. \label{eq: W}
\end{equation}

Substituting Eqs.~\eqref{eq: E} and \eqref{eq: W} into Jeffery's equation \eqref{eq:Jeffery} yields the angular velocities of the orientation angles:
\begin{eqnarray}
\dot{\chi}
&=& 
\frac{\partial \chi}{\partial t}
=
-\frac{1}{2} \dot \gamma \left[ 1 - \beta \cos (2\chi) \right], \label{eq:dot_chi}\\
\dot{\theta}
&=&
\frac{\partial \theta}{\partial t}
=
\frac{1}{4} \dot \gamma \beta \sin (2\chi) \sin(2\theta). \label{eq:dot_theta}
\end{eqnarray}
The explicit equations for a general flow field are provided in the Supplementary Information.

\subsection{Fokker-Planck equation}

We now derive the Fokker-Planck equation governing the orientational probability density of a rigid rod subjected to hydrodynamic and Brownian torques.

Let $\psi(\chi,\theta\,|\,t)$ be the orientation distribution function of a rod on the unit sphere.  
We consider the dilute regime in which there are no particle-particle interactions, so that rotational dynamics arise solely from hydrodynamic drift and rotational diffusion.
Under these conditions, probability conservation on the sphere yields \citep{leal1972,talbot2024}:
\begin{eqnarray}
    \frac{\partial \psi}{\partial t}
    =
    D_r \nabla_{\chi,\theta}^2 \psi
    -
    \nabla_{\chi,\theta}\cdot\big(\dot{\mathbf{\Omega}}\,\psi\big), \label{eq:FP_main}\\
    \int_{0}^{2\pi}\int_{0}^{\pi}\psi(\chi,\theta\,|\,t)\,\sin\theta\,d\theta\,d\chi=1. \label{eq:normalization}
\end{eqnarray}
Here $D_r$ is the rotational diffusion coefficient with units of $\mathrm{s}^{-1}$.
In the dilute limit, we use the slender-body approximation for the rotational diffusivity, where $D_r$ is independent of concentration and mainly depends on the rod length $l$ as \citep{tao2006,doi1978a}:
\begin{equation}
    D_r(l) =\frac{3 k_B T \ln(l/d)}{\pi \eta  l^3}, \label{eq:Dr_main}
\end{equation}
where $k_B$ is the Boltzmann constant, $T$ is the absolute temperature, and $\eta$ is the solvent viscosity.
The term $\dot{\mathbf{\Omega}}$ represents the deterministic angular drift induced by the imposed flow. 
The operators $\nabla_{\chi,\theta}^2$ and $\nabla_{\chi,\theta}\cdot$ denote, respectively, the Laplace-Beltrami operator and the surface divergence on the sphere:
\begin{eqnarray}
    \nabla_{\chi,\theta}^2 \psi
    =
    \frac{1}{\sin^2\theta}\frac{\partial^2 \psi}{\partial \chi^2}
    +
    \frac{1}{\sin \theta}\frac{\partial}{\partial \theta}\left(\sin \theta \frac{\partial \psi}{\partial \theta}\right), \label{eq:Laplace_operator}\\
    \nabla_{\chi,\theta}\cdot(\dot{\mathbf{\Omega}}\,\psi)
    =
    \frac{\partial}{\partial \chi}(\dot \chi \psi)
    +
    \frac{1}{\sin \theta} \frac{\partial}{\partial \theta}(\dot \theta\sin \theta  \psi). \label{eq:nabla_operator}
\end{eqnarray}

For simple shear flow, substituting Eqs.~\eqref{eq:dot_chi} and \eqref{eq:dot_theta} into Eq.~\eqref{eq:FP_main} gives:
\begin{equation}
    \frac{\partial \psi}{\partial t}
    =
    D_r \nabla_{\chi,\theta}^2 \psi
    -
    \dot \gamma \Lambda \psi, \label{eq:FP_simple_shear}
\end{equation}
where
\begin{equation}
\begin{split}
    \Lambda
    =
    \,-& \frac{3}{2} \beta \sin(2\chi)\sin^2\theta
    - \frac{1}{2}\left[1 - \beta \cos(2\chi)\right]\frac{\partial}{\partial \chi} \\
    +& \frac{1}{4}\beta \sin (2\chi)\sin(2\theta) \frac{\partial}{\partial \theta}.
    \end{split}
\end{equation}

To characterize the competition between drift and diffusion, 
we nondimensionalize time as $\tilde t=tD_r$, obtaining:
\begin{equation}
    \frac{\partial \psi}{\partial \tilde t}
    =
    \nabla_{\chi,\theta}^2 \psi
    -
    \Pe \Lambda \psi,
\end{equation}
where $\Pe = \dot \gamma / D_r$ is the P\'{e}clet number.

We solve Eq.~\eqref{eq:FP_simple_shear} numerically following \citet{doi1978}, which has also been adopted in other studies \citep{strand1987,talbot2024,rogers2005}.
In particular, we use a Galerkin spectral method based on real spherical harmonics.
These correspond to the exact solutions of the Laplace-Beltrami operator on the sphere, providing exact solutions when the drift is zero.
More details are provided in the Supplementary Information and in the code available online \citep{hillebrand2025}.

\subsection{Order parameter, orientation angle, and birefringence}

For a monodisperse system, the ensemble average with respect to $\psi$ is given by:
\begin{equation}\label{eq:ensemble_mono}
    \langle p \rangle
    =
    \int_{0}^{2\pi}\int_{0}^{\pi} p(\chi,\theta)\,\psi(\chi,\theta\,|\,t)\,\sin\theta\,d\theta\,d\chi.
\end{equation}
The in-plane scalar order parameter, which characterizes the degree of alignment relevant to flow birefringence, is defined as \citep{frattini1986,chow1985,fuller1995}:
\begin{eqnarray}\label{eq:order_mono}
    S
    =
    \sqrt{
    \langle q_x^2 - q_y^2 \rangle^2
    + \left( 2 \langle q_x q_y \rangle \right)^2
    }
    .
\end{eqnarray}
The corresponding orientation angle (also referred to as the extinction angle in flow birefringence measurements) is given by \citep{frattini1986,chow1985,fuller1995}:
\begin{equation}\label{eq:angle_mono}
    \phi
    =
    \frac{1}{2}\tan^{-1}\left(
    \frac{2 \langle q_x q_y \rangle}{\langle q_x^2 - q_y^2 \rangle}\right).
\end{equation}

The birefringence $\Delta n$ is directly proportional to the order parameter \citep{purdy2003,uetani2019,fuller1995}:
\begin{equation}
    \Delta n = \Delta n_{\max} S  , \label{eq:Delta_n_S}
\end{equation}
where $\Delta n_{\max}$ denotes the saturation birefringence for a fully aligned state under the given concentration.

\subsection{Polydispersity}

We consider suspensions in which rods are polydisperse in length while maintaining a fixed diameter. 
In contrast to the monodisperse case, the polydisperse orientation distribution function for dilute suspensions is described by the joint probability density function,
\begin{equation} \label{eq:pdf.poly}
    \psi_\mathrm{poly}(\chi, \theta, l \,|\, t) = \psi(\chi, \theta \,|\, t, l) f(l),
\end{equation}
where $\psi(\chi, \theta \,|\, t, l)$ is the monodisperse orientation distribution function conditioned on the rod length $l$ and described by Eq.~\eqref{eq:FP_simple_shear}, while $f(l)$ is some probability density function accounting for rod length polydispersity.
The ensemble average Eq.~\eqref{eq:ensemble_mono} is generalized to:
\begin{equation}\label{eq:ensemble_poly}
    \langle p \rangle_\mathrm{poly}
    =
    \int_0^\infty \int_{0}^{2\pi}\int_{0}^{\pi} p(\chi,\theta,l)\psi_\mathrm{poly}(\chi,\theta,l|t)\sin\theta d\theta d\chi dl.
\end{equation}
This is based on a similar concept to the method of previous studies \citep{hollingsworth1971,chow1985,rogers2005}.
The order parameter and orientation angle for a polydisperse system follow directly from Eq.~\eqref{eq:ensemble_poly} \citep{rogers2005}:
\begin{eqnarray}
    S
    &=&
    \sqrt{
    \langle q_x^2 - q_y^2 \rangle_\mathrm{poly}^2
    + \left( 2 \langle q_x q_y \rangle_\mathrm{poly} \right)^2
    } \label{eq:order_poly}\\
    \phi
    &=&
    \frac{1}{2}\tan^{-1}\left(
    \frac{2 \langle q_x q_y \rangle_\mathrm{poly}}{\langle q_x^2 - q_y^2 \rangle_\mathrm{poly}}\right). \label{eq:angle_poly}
\end{eqnarray}
Note that Eq.~\eqref{eq:Delta_n_S} does not strictly apply to polydisperse systems due to possible dependencies of material properties contained within $\Delta n_{\max}$ on $l$.
Here, we absorb such dependencies into the polydispersity distribution $f(l)$ instead, similar to other studies \citep{matsumoto1972,rogers2005,rogers2005a}.

While for monodisperse systems $f(l)$ can be considered a Dirac-delta function, it is not \textit{a priori} clear which polydispersity distribution is appropriate.
In polydisperse suspensions, the appropriate representation depends on (i) the measured number distribution of rod lengths and (ii) how material properties scale with rod length.
In this work, the physical rod population is characterized by the experimentally measured number distribution $f_0^\mathrm{exp}(l)$, as will be shown in Sec.~\ref{sec:experiment}, which we parametrize by a lognormal $f_0(l)$ to account for data sparsity in the long tail:
\begin{equation}\label{eq:lognormal}
    f_0(l)
    =
    \frac{1}{\sqrt{2 \pi }\sigma l} \exp \left[ - \frac{(\ln l - \mu)^2}{2 \sigma^2} \right],
\end{equation}
where $\mu$ is the logarithm of the median and $\sigma > 0$ is the logarithm of the geometric standard deviation.
Such lognormal forms are widely used for cellulose nanocrystals and related colloidal rods \citep{oakley1983,watson1992,jakubek2018}, and thus we consider it appropriate as a parametric model.

Because the experimentally observed optical response can be biased toward certain length classes through the aforementioned length-dependent contributions \citep{rogers2005, arenas-guerrero2018, matsumoto1972}, we define a moment-weighted lognormal family,
\begin{eqnarray}\label{eq:weighted_lognormal}
    f_n(l)
    &=&
    \frac{l^n f_0(l)}{\int_0^\infty l^n f_0(l) dl}\\ \nonumber
    &=&
    \frac{1}{\sqrt{2 \pi }\sigma l} \exp \left[ - \frac{(\ln l - (\mu + n\sigma^2))^2}{2 \sigma^2} \right],
\end{eqnarray}
where the weighting parameter $n$ controls the emphasis on longer rods.
Here, $n$ denotes the moment order, i.e., $f_0, f_1, f_2, f_3$ correspond to $n = 0, 1, 2, 3$, respectively.
This moment-weighted lognormal distribution $f_n(l)$ is used to calculate Eq.~\eqref{eq:pdf.poly}.
The corresponding mean length of $f_n(l)$ is:
\begin{equation}\label{eq:mean_length}
\langle l \rangle_n = \int_0^\infty l f_n(l) dl = \exp\left[\mu+\frac{(2n + 1)}{2}\sigma^2 \right].
\end{equation}
We then define an effective rotational diffusion coefficient as:
\begin{equation}\label{eq:Dreff}
\Dreff \equiv D_r(\langle l\rangle_n),
\end{equation}
and use $\Pe=\dot\gamma/\Dreff$ and $\tilde t=t\Dreff$ in our results.
Here, the effective rotational diffusivity $\Dreff$ is evaluated using the $n$-th moment-weighted average length $\langle l \rangle_n$ in Eq.~\eqref{eq:mean_length}.
We note that due to the special structure of the lognormal distribution, $f_n(l)$ are self-similar under the scaling $\tilde{l} = l / \langle l\rangle_n$ with fixed $\sigma$.
As we will show, this causes the weighting parameter $n$ to control primarily a horizontal shift with respect to $\dot\gamma$ but not the shape of the predicted values, which is instead controlled by $\sigma$.

\subsection{Relaxation dynamics after flow cessation}

Under steady shear, the hydrodynamic torque imposed by the velocity gradient competes with rotational diffusion, producing a non-isotropic orientation distribution $\psi(\chi,\theta)$ and a finite scalar order parameter $S$, measurable as birefringence $\Delta n/\Delta n_\mathrm{max}$, see Eq.~\eqref{eq:Delta_n_S}.
When the shear is suddenly stopped at $t=0$, the hydrodynamic alignment torque vanishes and the subsequent evolution is governed solely by rotational diffusion, causing both $\psi(\chi,\theta\,|\,t)$ and $\Delta n(t)$ to relax toward their isotropic equilibrium states.

For a monodisperse system of rods with identical length $l$, the relaxation of the order parameter $S(t)$ follows a single exponential decay characterized by the orientation relaxation time $\tau_b = 1/6D_r$ (see also the Supplementary Information).
Because this timescale depends only on the rotational diffusivity, the decay of $S(t)$ is independent of the shear rate applied prior to cessation.
Explicitly, the order parameter decays as:
\begin{equation}\label{eq:decay_mono}
    S(t) = S|_{t = 0} \exp \left(-t/\tau_b\right),
\end{equation}
where $S|_{t = 0}$ is the steady-state value before the shear is stopped.

In contrast, polydisperse suspensions exhibit intrinsically a multi-mode relaxation because each rod length $l$ relaxes with its own characteristic time $\tau_b(l)=1/6D_r(l)$ and the relation to $S$ and $\phi$ are generally nonlinear.
To characterize the overall relaxation behavior in polydisperse systems with a single experimentally accessible quantity, we define  the ideal averaged relaxation time as \citep{matsumoto1972}:
\begin{equation}\label{eq:taub_poly}
    \bar{\tau}_b
    =
    \int_0^{\infty} \frac{S(t)}{S|_{t = 0}} dt
    =
    \int_0^{\infty} \frac{\Delta n(t)}{\Delta n|_{t = 0}} dt.
\end{equation}
This definition provides a single measurable timescale that captures the overall relaxation behavior, while still embedding the influence of the underlying rod length polydispersity and the initial alignment state.
Furthermore, substituting Eq.~\eqref{eq:decay_mono} into Eq.~\eqref{eq:taub_poly} yields the consistent result $\bar{\tau}_b = \tau_b$ for monodisperse systems.
In the experiment, the observation window is finite and affected by experimental noise.
Therefore, the integral is performed only until an end of observation time $t_e=3/\Dreff$.
This will result in an underestimation of $\bar{\tau}_b$ but allows for a direct comparison with experiments.

\section{Experiment}
\label{sec:experiment}

In this section, we describe the materials, experimental setup, and measurement protocols used to probe steady-state alignment and relaxation after flow cessation in dilute CNC suspensions.

\subsection{Working fluids}

\begin{figure*}[t]
    \centering
    \includegraphics[width=2\columnwidth]{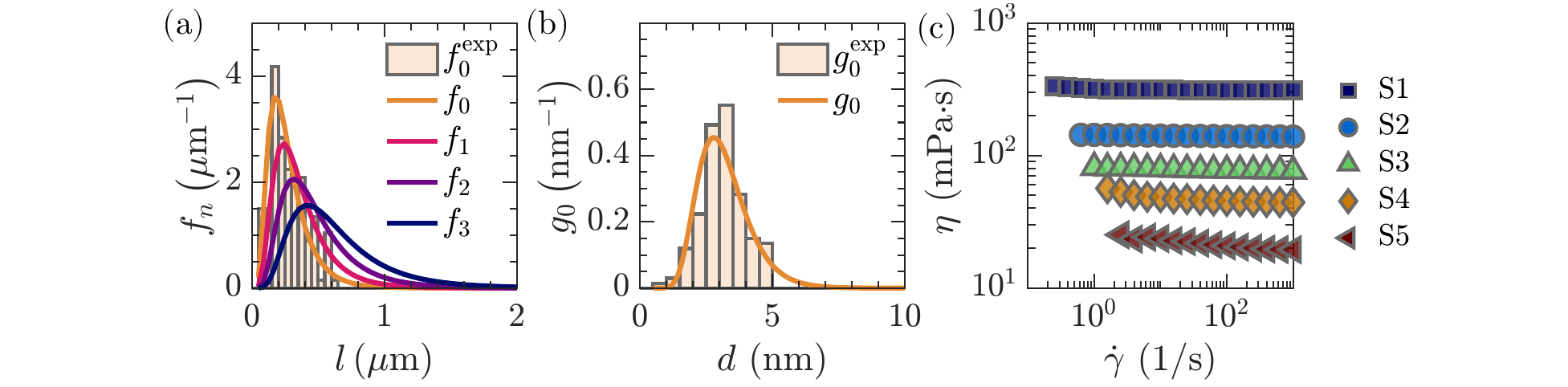}
    \caption{\label{fig:Viscosity_LengthDistribution}
    Characterization of the CNC suspensions used in the experiments. (a) Rod-length distributions $f_n(l)$ and (b) rod-diameter distribution $g_0(d)$ of CNCs. 
    The histograms represent number-normalized distributions $f_0^\mathrm{exp}(l)$ and $g_0^\mathrm{exp}(d)$ extracted from AFM imaging.
    The solid curve for $f_0(l)$ is a lognormal fit to $f_0^\mathrm{exp}(l)$, while $f_n(l)\propto l^n f_0(l)$ ($n=0$-$3$) is a moment-weighted family derived from the same fitted $f_0(l)$.
    The fitted parameters for $f_0(l)$ are $\exp(\mu) = 238.4\,\mathrm{nm}$ and $\exp(\sigma^2) = 1.3275$.
    For the diameter distribution, the mean diameter $\langle d\rangle_0 = 3.2\,\mathrm{nm}$ is used as a constant in the theoretical calculations.
    (c) Shear viscosity $\eta$ of CNC suspensions at different concentrations ($0.005$-$0.20$\,wt\%) in glycerol-water mixtures, measured with a cone-plate rheometer at 25$\rm ^\circ C$.}
\end{figure*}

\begin{table*}[]
    \centering
    \caption{\label{tab:spec_suspensions}
        Summary of the CNC suspensions used in the experiments. 
        }
    \begin{tabular}{cccccccc}
    \hline
    Label
    & $c_{m,G} ~\mathrm{(wt\%)}$
    & $c_m ~\mathrm{(wt\%)}$
    & $\rho ~\mathrm{(g/cm^3)}$
    & $c_v ~\mathrm{(vol\%)}$
    & $\kappa l^3$ & $\eta ~\mathrm{(mPa\cdot s)}$
    & $\tau_f ~\mathrm{(ms)}$ \\ \hline
    S1 & 95 & 0.005 & 1.24 & 0.0042 & 0.39 & 312.2 & 1.1  \\
    S2 & 90 & 0.02  & 1.23 & 0.016  & 1.5  & 139.0 & 2.3  \\
    S3 & 85 & 0.05  & 1.21 & 0.040  & 3.8  & 76.5  & 4.2  \\
    S4 & 80 & 0.1   & 1.20 & 0.080  & 7.5  & 44.4  & 7.2  \\
    S5 & 70 & 0.2   & 1.17 & 0.156  & 15   & 19.5  & 16.3 \\ \hline
    \end{tabular}
\end{table*}

CNC suspensions were prepared by dispersing CNCs (CelluForce, 5.6\,wt\% aqueous stock suspension) at different concentrations in glycerol-water mixtures.
The morphology of the CNCs was characterized from atomic force microscopy (AFM) images reported previously by \citet{calabrese2024,calabrese2021}. 
Figures~\ref{fig:Viscosity_LengthDistribution}(a) and (b) show the number-normalized length and diameter distributions $f_0^\mathrm{exp}(l)$ and $g_0^\mathrm{exp}(d)$, respectively, extracted from these AFM images using FiberApp \citep{usov2015}. 
The CNCs typically have lengths $l$ of $100$-$500\,\mathrm{nm}$ and diameters $d$ of $2$-$5\,\mathrm{nm}$, yielding aspect ratios $r_p = l/d$ in the range $20$-$250$.
This broad distribution of rod sizes produces a wide range of rotational diffusion coefficients $D_r(l)$, which strongly influence orientation relaxation. 

The solid lines in Figs.~\ref{fig:Viscosity_LengthDistribution}(a) and (b) show lognormal distributions given by Eq.~\eqref{eq:lognormal} fitted to $f_0^\mathrm{exp}(l)$ and $g_0^\mathrm{exp}(d)$. 
For the length distribution $f_0(l)$, the fit yields $\exp(\mu) = 238.4\,\mathrm{nm}$ and $\exp(\sigma^2) = 1.3275$, whereas for the diameter distribution $g_0(d)$ we obtain $\exp(\mu) = 3.1\,\mathrm{nm}$ and $\exp(\sigma^2) = 1.0947$.
Increasing the weighting parameter $n$ gradually shifts the fitted curves toward longer lengths, increasing the median length $\exp(\mu)$ and the mean length \eqref{eq:mean_length} by a factor $\exp(\sigma^2)^n$, thus indicating that longer rods contribute more strongly to the weighted distributions.
Using the fitted lognormal functions $f_0(l)$ together with Eq.~\eqref{eq:mean_length}, we obtain the weighted mean rod lengths $\langle l \rangle_n$ for each $n$ as
$\langle l \rangle_0 = 274.7\,\mathrm{nm}$,
$\langle l \rangle_1 = 364.7\,\mathrm{nm}$,
$\langle l \rangle_2 = 484.1\,\mathrm{nm}$, and
$\langle l \rangle_3 = 642.6\,\mathrm{nm}$.
These weighted mean lengths are used to calculate the corresponding effective rotational diffusion coefficients $\Dreff$, which serve as the characteristic diffusion coefficient when comparing the experimental data. 

The rod diameter can be treated as effectively constant in the present analysis since its polydispersity is small compared to that of length.
Using the fitted diameter distribution $g_0(d)$ and Eq.~\eqref{eq:mean_length} with $l$ replaced by $d$, we obtain a mean diameter of $\langle d \rangle_0 = 3.2\,\mathrm{nm}$.
We use this value as a fixed rod diameter in the calculations of both the monodisperse and polydisperse Fokker-Planck equations.

CNC suspensions were prepared at 5 different concentrations expressed in terms of the CNC mass fraction $c_m$.
For quantitative comparison with theory, these concentrations were converted to the CNC volume fraction $c_v$ using the CNC density $\rho_p = 1500\,\mathrm{kg/m^3}$ \citep{calabrese2024} and the densities of glycerol (Sigma-Aldrich, 99\%) and water (Milli-Q, Millipore), $\rho_G = 1260\,\mathrm{kg/m^3}$ and $\rho_W = 997\,\mathrm{kg/m^3}$, respectively.
The glycerol mass fraction $c_{m,G}$ and the CNC mass fraction $c_m$ of each sample, together with the corresponding value of $c_v$ and the resulting suspension density $\rho$, are summarized in Table~\ref{tab:spec_suspensions}.

Here, we evaluate the concentration of the CNC suspensions.
To quantify the concentration regime, we calculate the dimensionless number density parameter $\kappa l^3 = 4 c_v l^2/(\pi d^2)$, where $d$ and $l$ are the rod diameter and representative length, and $c_v$ is the rod volume fraction.
The volume fraction is defined as $c_v = \rho c_m/\rho_p$. 
The values of $\kappa l^3$ for each suspension are listed in Table~\ref{tab:spec_suspensions}. 
Here, we used $\langle l \rangle_0 = 274.7 \, \mathrm{nm}$ to calculate $\kappa l^3$.
For rod-like colloids, the dilute regime corresponds to $\kappa l^3 \lesssim O(1)$ \citep{doi1978a,doi1988}, whereas excluded-volume interactions become significant and corrections to $D_r$ are required once $\kappa l^3 \gtrsim 10$ \citep{tao2006,oh1992}. 
According to this criterion, samples S1-S3 lie in the dilute regime, whereas S4 and S5 approach the onset of the semi-dilute regime. 
\citet{bertsch2019} performed small-angle X-ray scattering on CNCs identical to our source, indicating interaction onset at 0.5\,wt\%.
Previously, \citet{calabrese2022} demonstrated that concentration-dependent rotational diffusion emerges above 0.2\,wt\% due to rod interaction effects.
At lower levels (0.1\,wt\%, 0.2\,wt\% here), birefringence and orientation data collapse using the rotational diffusion coefficient from the dilute regime.
Since all CNC concentrations are 0.2\,wt\% or less in this work, we use the dilute-limit expression for $D_r$ for all samples.
As will be shown in Sec.~\ref{sec:RandD_Steady}, data collapse in concentration-normalized birefringence also justifies using the dilute-limit expression for $D_r$.

The solvent viscosity $\eta$, which sets both the rotational P\'{e}clet number and the flow-relaxation time, was controlled by varying the glycerol concentration in the glycerol-water mixtures (Table~\ref{tab:spec_suspensions}).
The shear viscosity of each sample was measured using a cone-plate rheometer (Anton Paar, MCR302 with a 50\,mm diameter cone plate) over the range $0.1 \leq \dot \gamma \leq 10^3\,\mathrm{s^{-1}}$.
The resulting viscosity curves, shown in Fig.~\ref{fig:Viscosity_LengthDistribution}(c), exhibit nearly Newtonian behavior at low $c_v$ and weak shear thinning at higher $c_v$.
For comparison with the theory, we treat all suspensions as Newtonian and use the viscosities $\eta$ at $\dot \gamma = 10^3\,\mathrm{s^{-1}}$, summarized in Table~\ref{tab:spec_suspensions}.
Based on the CNC concentration and the dilute assumption, the measured viscosities at high-shear rate of each sample are assumed to be nearly identical to their solvent viscosities.

\subsection{Experimental setup}

\begin{figure*}[t]
    \centering
    \includegraphics[width=1.9\columnwidth]{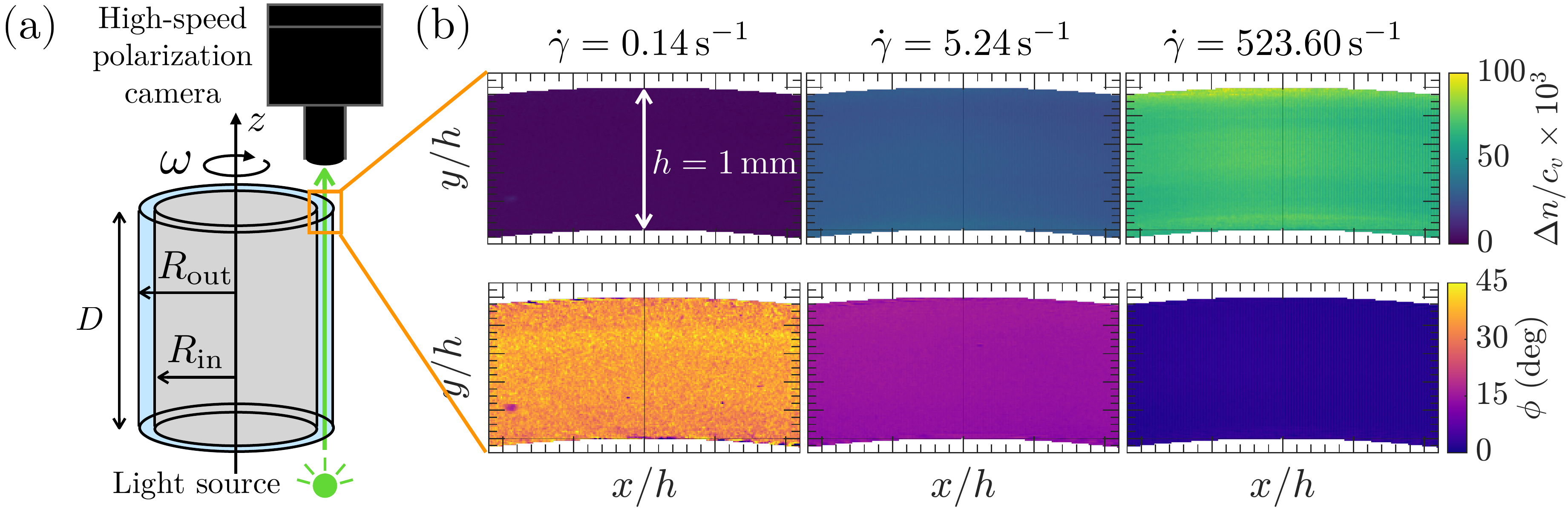}
    \caption{\label{fig:setup}
    (a) Schematic of the Taylor-Couette/polarization-imaging setup. 
    (b) Typical fields of concentration-normalized birefringence $\Delta n/c_v$ and orientation angle $\phi$ at different shear rates $\dot \gamma$ (Sample S1), where $c_v$ denotes the CNC volume fraction (dimensionless).
    The birefringence is evaluated as $\Delta n=\delta/D$, where $\delta$ is the measured retardation and $D$ is the optical path length, and $\phi$ is measured relative to the $x$-axis.
    The approximately uniform fields across the gap justify the use of gap-averaged quantities in comparison with theory.}
\end{figure*}

To impose a well-controlled simple shear flow, we used a narrow-gap Taylor-Couette cell that closely approximates planar Couette flow. 
A schematic diagram of the experimental setup is shown in Fig.~\ref{fig:setup}(a). 
The rotatable inner cylinder had radius $R_\mathrm{in} = 10\,\mathrm{mm}$ while the fixed outer cylinder had radius $R_\mathrm{out} = 11\,\mathrm{mm}$ giving a gap width $h = R_\mathrm{out} - R_\mathrm{in} = 1\,\mathrm{mm}$.
The height of the inner cylinder was $D = 34\,\mathrm{mm}$.
Both cylinders were custom-machined from aluminum using a computer numerical control lathe to ensure concentric alignment.
Rotation of the cylinder was driven by a NEMA 8 stepper motor, either directly or via a gearbox, depending on the required range of angular velocity $\omega$. 
A geared stepper motor with a planetary gearbox (8HS15-0604S-PG90, gear ratio 90:1) was used for steady and low-speed operation, providing high torque and precise control of the angular velocity. 
A direct-drive stepper motor without a gearbox (8HS15-0604S) was employed for higher-speed operation. 
Both configurations enabled stable rotation of the inner cylinder with minimal vibration and accurate control of the shear rate.
Two glass windows at the top and bottom provided optical access along the vertical axis.
This large height, corresponding to the optical path for the birefringence experiments, enables measurements of rod alignment at low concentrations. 

The narrow-gap configuration ($h/R_\mathrm{in} = 0.1$) ensures that the velocity profile across the gap is nearly linear, so that the approximated that of a Couette. 
The nominal shear rate was therefore estimated as $\dot \gamma = U/h$, where $U = \omega R_\mathrm{in}$ is the inner-wall velocity.
From the exact Taylor-Couette solution, the shear rate at the gap midpoint $r = (R_\mathrm{in} + R_\mathrm{out})/2$ is $\dot \gamma_\mathrm{mid} = 8 \omega R_\mathrm{in}^2 R_\mathrm{out}^2 /h(R_\mathrm{out} + R_\mathrm{in})^3$, which yields $\dot \gamma_\mathrm{mid} = 1.045\, U/h$ for the present geometry \citep{papanastasiou2021}.
Thus, the planar Couette approximation $\dot \gamma = U/h$ underestimates the exact shear rate by only $4.5\%$ and is sufficiently accurate for the present analysis.

For planar Couette flow between parallel plates separated by $h$, the unsteady Stokes equation with no-slip boundary conditions yields an exponential decay of the shear rate with a characteristic time
\begin{equation}
    \tau_f = \frac{h^2}{\pi^2 \nu},
\end{equation}
where $\nu$ is the kinematic viscosity of the suspending liquid (see the Supplementary Information for the derivation).
The values of $\tau_f$ for each sample are summarized in Table~\ref{tab:spec_suspensions}. 
Because $\tau_f$ decreases with increasing viscosity, the highly viscous glycerol-water mixtures used in this study lead to very rapid decay of the shear rate once the inner cylinder is stopped. 
As will be shown in Sec.~\ref{sec:RandD_Relaxation}, under the present experimental conditions, $\tau_f$ is much smaller than the orientation relaxation time $\bar{\tau}_b$, so the flow can be regarded as being switched off quasi-instantaneously at $t=0$.
The subsequent decay of birefringence therefore directly reflects orientation relaxation under effectively stopped-flow conditions.
We also tested the flow (not rod orientation) relaxation behavior of S5 via micro PIV and confirmed that the theoretical prediction, $\tau_f = 16.3\,\mathrm{ms}$, reproduced the actual relaxation time of the flow reasonably well.
We note that, in a regime where $\bar{\tau}_b$ becomes comparable to $\tau_f$, the measured birefringence decay would reflect a convolution of orientational relaxation with the finite hydrodynamic shutoff $\dot\gamma(t)$ rather than an instantaneous cessation.
The present theoretical comparisons therefore assume instantaneous cessation and are expected to be most accurate in the experimentally accessed regime where $\tau_f \ll \bar{\tau}_b$.

To measure flow-induced birefringence and orientation angle, the sample was illuminated from below with an intense LED light source (REVOX, SLG-150V) equipped with a band-pass filter (wavelength $\lambda = 520\,\mathrm{nm}$) and a circular polarizer.
The transmitted light passed through the sheared CNC suspension and was captured using a high-speed polarization camera (Photron, CRYSTA PI-1P). 
Each super-pixel of the camera consists of four sub-pixels with linear polarizers at $0^\circ$, $45^\circ$, $90^\circ$, and $135^\circ$, allowing simultaneous capture of the four intensity components required for birefringence analysis. 
The retardation $\delta$ and the orientation angle $\phi$ (measured relative to $x$-axis) were calculated from the four intensity components using the four-step phase-shifting method, as described in previous studies \citep{onuma2014,yokoyama2023}.
These quantities were obtained using commercial software (Photron, CRYSTA Stress Viewer) based on the intensity images recorded by the polarization camera. 
The effective spatial resolution of the processed retardation field was one quarter of the raw image pixel resolution, corresponding to approximately $10\,\mathrm{\mu m/pix}$.
The temporal resolution was up to $250 \,\mathrm{fps}$, which is sufficient to capture the unsteady birefringence during flow cessation.
All experiments were conducted at room temperature, 25 $\rm ^\circ C$.

Before measurement, the software subtracted background images of quiescent suspensions at rest.
Assuming a two-dimensional flow field that is uniform along the optical axis ($\Delta n(z) = \mathrm{const.}$), the birefringence was obtained as $\Delta n = \delta/D$, where $D$ is the optical path length (height of the inner cylinder).
Figure~\ref{fig:setup}(b) shows examples of $\Delta n/c_v$ and $\phi$ fields measured by the camera at different $\dot \gamma$.
The fields exhibit uniform distribution across the gap and ensure the Couette flow assumption.
For the analysis, we focus on the birefringence and orientation angle averaged over the gap between inner and outer cylinders.
For steady shear conditions, the data shown correspond to values that are averaged over a sufficiently developed steady-state interval of at least $2\,\mathrm{s}$ prior to flow cessation.
The error bars represent the standard deviation of these quantities over this averaging window.
In the experiment, birefringence data were collected three times for the same shear rate. 
Unless otherwise noted, all three sets of data points are displayed in the figures.

To probe the relaxation dynamics, the shear flow was abruptly stopped at $t = 0$. 
The subsequent decay of birefringence was recorded up to a final time $t_e = 3/\Dreff$ to capture several characteristic relaxation times and ensure that the birefringence signal decayed to a small fraction of its initial value.
Here, $\Dreff$ depends on the solvent viscosity $\eta$ and a parameter $n$ determining the weighted mean length of the rods $\langle l \rangle_n$.
$\Dreff$ is computed from Eq.~\eqref{eq:Dreff}, using the prescribed length distribution, and is not obtained from the experimentally measured birefringence relaxation time $\bar{\tau}_b$.
For the monodisperse case, this corresponds to an observation window of $t_e \approx 18\tau_b$ based on $\tau_b=1/(6D_r)$.
The averaged relaxation time $\bar{\tau}_b$ defined in Eq.~\eqref{eq:taub_poly} is evaluated from the measured $\Delta n(t)$ by numerical integration up to $t_e$.

The steady-state values of $\Delta n$ and $\phi$, as well as the relaxation curves after flow cessation obtained in this way, are analyzed and compared with the theoretical predictions in the next section.

\section{Results and discussion}
\label{sec:results}

We characterize steady-state alignment of dilute mono- and polydisperse rod suspensions under simple shear using the Fokker-Planck model and compare the predictions with experimentally measured flow birefringence and orientation angle.
We then analyze relaxation after flow cessation, quantify the dependence of the averaged relaxation time on the pre-shear rate, and compare the results with the polydisperse model.

\subsection{Steady-state alignment: theory and experiment}\label{sec:RandD_Steady}

\begin{figure*}[t]
    \centering
    \includegraphics[width=2\columnwidth]{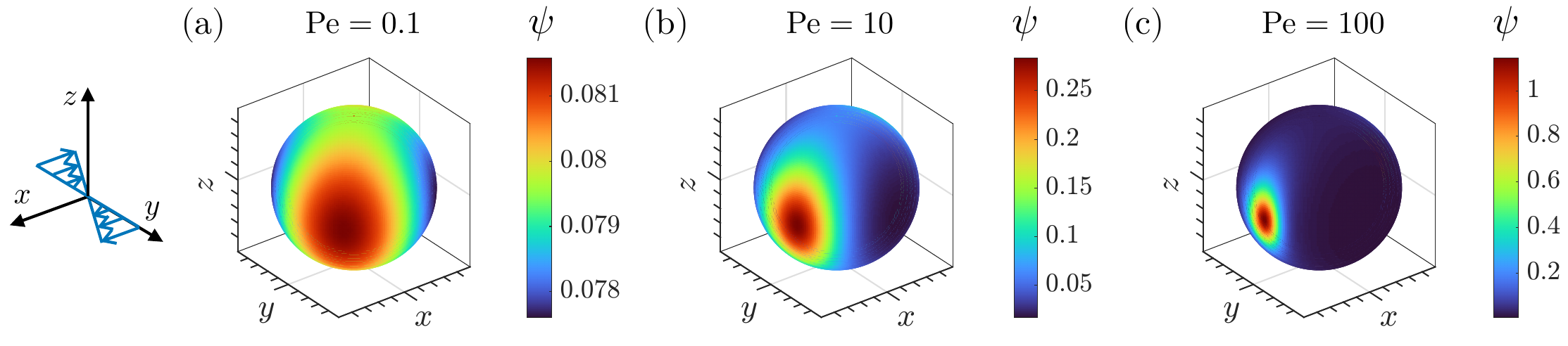}
    \caption{
    Steady-state orientation distribution function $\psi(\chi,\theta)$ predicted for a monodisperse suspension at different P\'{e}clet numbers with constant diameter $d=3.2\,\mathrm{nm}$:
    (a) $\Pe = 0.1$,
    (b) $\Pe = 10$, and
    (c) $\Pe = 100$.}
    \label{fig:Psi_Steady}
\end{figure*}

Figure~\ref{fig:Psi_Steady} shows examples of the steady-state orientation distribution function $\psi(\chi,\theta)$ for a monodisperse suspension at different $\Pe$. 
As $\Pe$ increases, the distribution evolves from a nearly isotropic state to a strongly anisotropic state with a pronounced peak near the flow direction, reflecting the increasing dominance of shear-induced alignment over rotational diffusion.
Specifically, at $\Pe = 0.1$, the distribution remains nearly isotropic with a very narrow range of $\psi(\chi,\theta)$, indicating that rotational diffusion dominates over flow-induced alignment.
Only a weak maximum of $\psi(\chi,\theta)$ appears at $\theta = 90^\circ$ (within the $xy$-plane) and $\chi \approx 45^\circ$ (and $225^\circ$).
These directions coincide with the principal axes of the strain-rate tensor $\mathbf{E}$, along which the rods experience the largest extensional and compressional components of the flow.
At $\Pe = 10$, the distribution becomes markedly anisotropic, with a narrow maximum near the flow direction ($\chi \approx 0^\circ$ and $180^\circ$).
At $\Pe = 100$, $\psi(\chi,\theta)$ exhibits a sharp peak, indicating strong alignment along the flow direction.
This transition from nearly isotropic to highly aligned distributions reflects the increasing dominance of flow-induced torques over rotational diffusion as $\Pe$ increases.

\begin{figure*}[t]
    \centering
    \includegraphics[width=2\columnwidth]{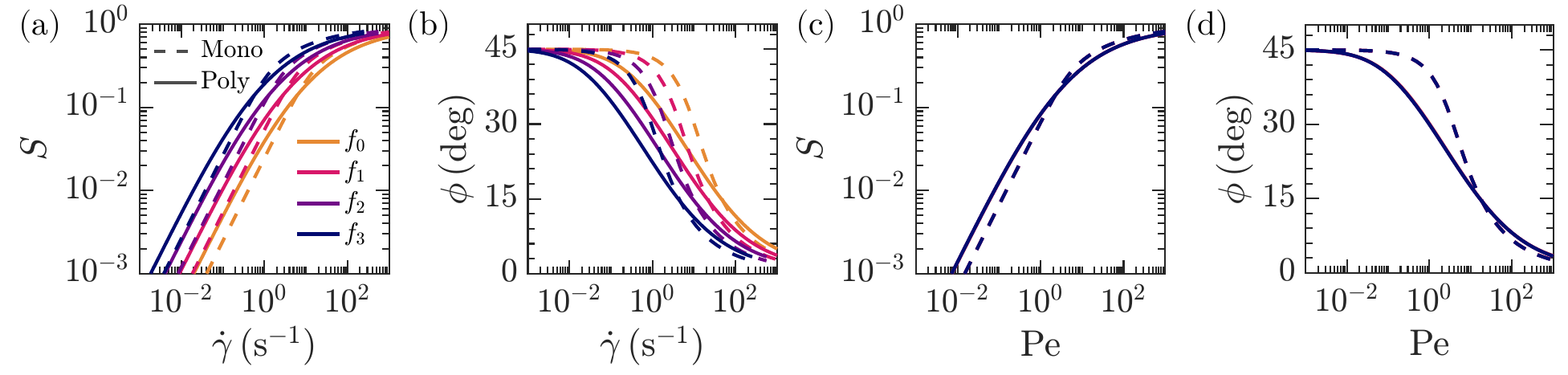}
    \caption{
    (a) \& (b) Steady-state order parameter $S$ and orientation angle $\phi$ as functions of $\dot \gamma$, and (c) \& (d) the same quantities as functions of $\Pe$, for different $n$ and constant diameter $d = 3.2\,\mathrm{nm}$.
    Dashed and solid lines represent the monodisperse and polydisperse cases, respectively.
    $f_n$ denotes the $n$-th moment-weighted distribution (e.g., $f_0, f_1, f_2,f_3$ for $n = 0, 1, 2, 3$).
    We define $\Pe = \dot \gamma/\Dreff$ using an effective rotational diffusion coefficient $\Dreff = D_r(\langle l \rangle_n)$ based on the weighted mean length $\langle l \rangle_n$.
    The polydispersity distributions $f_n(l)$ are based on the fitted number-weighted distribution of CNC rods (see Fig.~\ref{fig:Viscosity_LengthDistribution}(a)), yielding $\exp(\mu) = 238.4\,\mathrm{nm}$ and $\exp(\sigma^2) = 1.3275$.
    Here, for illustration we evaluate $\Dreff$ using the viscosity of sample S1; the resulting values are
    $\Dreff = 2.7, 1.2, 0.56,$ and $0.25\,\mathrm{s^{-1}}$ for $n=0,1,2,$ and $3$, respectively.
    }
    \label{fig:SvsPe}
\end{figure*}

\begin{figure*}
    \centering
    \includegraphics[width=2\columnwidth]{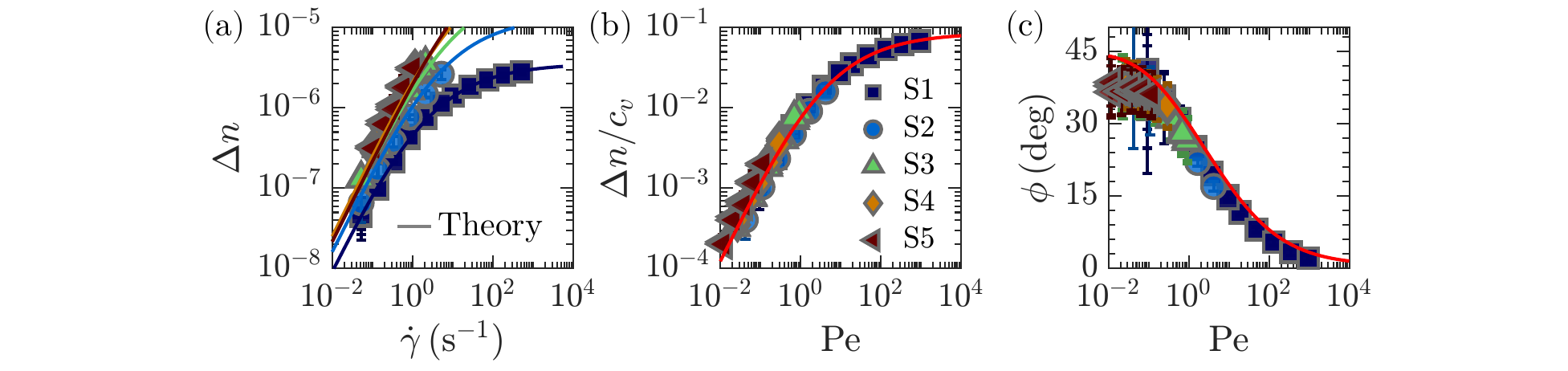}
    \caption{
    Steady-state flow birefringence $\Delta n$ and orientation angle $\phi$ of CNC suspensions under simple shear. 
    (a) Birefringence $\Delta n$ versus shear rate $\dot \gamma$. 
    (b) Concentration-normalized birefringence $\Delta n/c_v$, with the CNC volume fraction $c_v$ (dimensionless) and (c) orientation angle $\phi$ versus the P\'{e}clet number $\Pe = \dot \gamma/\Dreff$.
    Here, $\Dreff$ is computed as $D_r(\langle l\rangle_2)$ using Eq.~\eqref{eq:Dr_main}, where $n=2$ is chosen due to the overall best fit between theory and experiment.
    Symbols denote experimental data at different CNC concentrations and solvent viscosities. 
    The error bars indicate the standard deviation over the averaging time window.
    Solid curves represent steady-state predictions of the polydisperse Fokker-Planck model evaluated with the experimentally measured $f_0(l)$ and fixed $n=2$.
    To compare theory with experiment, we use $\Delta n_\mathrm{max}/c_v = 9 \times 10^{-2}$ in Eq.~\eqref{eq:Delta_n_S} based on a simple fit with experimental data.}
    \label{fig:BvsPe}
\end{figure*}

From $\psi(\chi,\theta)$, the scalar order parameter $S$ and orientation angle \(\phi\), which quantitatively describe the narrowing of the peak and its shift toward the flow direction, are computed using Eqs.~\eqref{eq:order_mono}-\eqref{eq:angle_mono} for monodisperse and
Eqs.~\eqref{eq:order_poly}-\eqref{eq:angle_poly} for polydisperse predictions.
Figure~\ref{fig:SvsPe}(a) and (b) show $S$ and $\phi$ as functions of the applied shear rate $\dot \gamma$ for different weighting parameters $n$.
The dashed lines and solid lines correspond to the monodisperse and polydisperse cases, respectively, and the colors indicate the different $n$.
For each $n$, the monodisperse case is calculated using rods of length $\langle l \rangle_n$, consistent with the definition of the effective rotational diffusion coefficient $\Dreff$.

As $\dot \gamma$ increases, $S$ increases from a diffusion-dominated regime (weak alignment) to a flow-dominated regime where $S$ approaches a plateau close to unity, reflecting strong streamwise alignment.
This trend is observed in both monodisperse and polydisperse cases for all weighting parameters, see Fig.~\ref{fig:SvsPe}(a).
At a fixed shear rate, the polydisperse predictions exhibit a systematic dependence on $n$.
Increasing $n$ assigns greater statistical weight to longer rods, which more readily align under shear, thereby yielding larger $S$.
Comparison between monodisperse and polydisperse cases further highlights the role of the length distribution.
At relatively low $\dot \gamma$, the polydisperse $S$ is larger than the monodisperse case because the long-rod tail contributes a fraction of strongly aligned particles even when the mean-length monodisperse system remains only weakly aligned.
At sufficiently high $\dot \gamma$, this trend reverses.
Shorter rods within the distribution, having larger rotational diffusivity and thus weaker alignment at the same $\dot \gamma$, reduce the ensemble-averaged order parameter, leading to polydisperse values of $S$ that are slightly smaller than the monodisperse case.

Similar observations apply to the orientation angle $\phi$ in Fig.~\ref{fig:SvsPe}(b).
In the diffusion-dominated regime at low $\dot \gamma$, the orientational distribution is broad and the mean orientation approaches the principal axis of $\mathbf{E}$, i.e., $\phi \approx 45^\circ$.
With increasing $\dot \gamma$, flow-induced alignment strengthens and $\phi$ decreases toward $0^\circ$, consistent with preferential alignment along the flow direction.
The dependence on $n$ mirrors that of $S$: larger $n$ (greater emphasis on longer rods) yields smaller $\phi$ at a given $\dot \gamma$ due to overall stronger streamwise alignment.
Relative to the monodisperse case, polydispersity produces slightly smaller $\phi$ at low $\dot \gamma$ and slightly larger $\phi$ at high $\dot \gamma$ following the same mechanism as described for $S$.

Figure~\ref{fig:SvsPe}(c) and (d) show $S$ and $\phi$ compared with $\Pe=\dot \gamma/\Dreff$, where $\Dreff=D_r(\langle l\rangle_n)$.
The strong $n$-dependence seen in panels (a) and (b) largely disappears and the curves collapse onto master curves for $S$ and $\phi$ in both the monodisperse and polydisperse cases.
This collapse is a direct consequence of the self-similar structure of the weighted lognormal distributions under the transformation $\tilde{l} = l / \langle l \rangle_n$ and the relatively close $\beta \lesssim 1$ values.
Consistent with this invariance, the dominant effect of changing $n$ is absorbed into $\Dreff$ through $\langle l\rangle_n$.
In other words, increasing $n$ shifts the $\dot \gamma$-based curves horizontally, compare for example panels (a) and (c).
In experiments, the measured steady birefringence is controlled only by the dimensional shear rate $\dot\gamma$ for a given suspension.
The parameter $n$ cannot change the experimental data; it only changes the definition of the effective diffusion coefficient $\Dreff$ used for $\Pe=\dot\gamma/\Dreff$, and thus controls the horizontal shift of experimental points of birefringence and orientation data when plotted against $\Pe$.
We use this observation to identify $n$ in the modeled polydispersity distribution based on experimental data.

To further clarify the influence of polydispersity (effect of $\sigma$ of length distribution), Fig.~S.2(a-c) in the Supplementary Information presents the theoretical predictions obtained by varying $\sigma$ of the lognormal length distribution, where the self-similarity no longer holds.
Increasing $\sigma$ enhances the contribution of long rods and yields larger steady-state order parameters and smaller extinction angles in the low-$\Pe$ regime.
These trends are consistent with the stronger alignment tendency of slowly diffusing long rods and complement the theoretical trend in Fig.~\ref{fig:SvsPe}.

Figure~\ref{fig:BvsPe} summarizes the experimental steady-state flow birefringence $\Delta n$ and orientation angle $\phi$ (symbols) compared with theory (solid lines). 
Panel (a) shows the measured birefringence $\Delta n$ against the applied shear rate $\dot \gamma$, while panels (b) and (c) plot the concentration-normalized birefringence $\Delta n/c_v$ and the orientation angle $\phi$ against $\Pe = \dot \gamma/ \Dreff$, respectively.
We use the average length $\langle l \rangle_2$ (with $n=2$) to compute $\Dreff$ because, as we will discuss shortly, it provides the best overall fit between theory and experiment, see also solid lines in Fig.~\ref{fig:BvsPe} (for $n=0,1,3$ see Fig.~S.3 in the Supplementary Information for a comparison).
To obtain the theoretical predictions for $\Delta n$, we fit the calculated $S$ to experimental $\Delta n/c_v$ resulting in $\Delta n_\mathrm{max}/c_v = 9 \times 10^{-2}$.
This lies within the range 0.0057-0.12 (0.00597 $\pm$ 0.0230, mean $\pm$ standard deviation) previously reported for CNCs \citep{khan2023}.

In the low-$\Pe$ regime ($\Pe < 1$), see Fig.~\ref{fig:BvsPe}(b), $\Delta n/c_v$ increases linearly with $\Pe$, consistent with the theoretical prediction $S \propto \Pe$, see solid red line.
Around $\Pe \sim 1$, the slope of $\Delta n/c_v$ begins to decrease, corresponding to the onset of significant flow alignment.
For $\Pe \gg 1$, $\Delta n/c_v$ approaches a plateau, indicating near-complete alignment of the rods.
Figure~\ref{fig:BvsPe}(c) shows that the orientation angle $\phi$ decreases monotonically with increasing $\Pe$, transitioning from $\phi \approx 45^\circ$ in the Brownian-dominated regime to values approaching $0^\circ$ under strong flow alignment, again consistent with theory.
In the low-$\Pe$ regime, the experimental uncertainty of $\phi$ is relatively large, as reflected by the error bars, due to the weak birefringence signal and the almost isotropic orientation distribution.
This is reflected when comparing with theory, where the predicted orientation angle appears 2-3$^\circ$ larger.
Nevertheless, we see an excellent match between theory and experiment.

Across all concentrations, the experimental data collapse when plotted against $\Pe$ and hence demonstrates that $\dot \gamma/\Dreff$ using an average length $\langle l \rangle_2$ is an appropriate scaling parameter to capture the competition between shear-induced alignment and Brownian rotational diffusion in polydisperse rod suspensions.
Beyond supporting the dilute-limit assumption, this collapse is practically significant: it indicates that a single properly defined P\'{e}clet number can serve as a predictive control parameter for flow-induced alignment across a range of CNC concentrations, despite the underlying polydispersity.
The measurements span more than five orders of magnitude in $\Pe$, from the diffusion-dominated regime to the flow-aligned regime.

It is useful to compare the continuous polydispersity addressed in this study with ``bidisperse suspensions,'' in which the effects are separated into two distinct length classes.
The study by \citet{lang2020} demonstrated that the rheological behavior of such bidisperse rod suspensions is governed by the coexistence of two characteristic rotational diffusion times associated with the short and long rods.
We note, however, that their experiments were performed in the semidilute regime, where interparticle interactions influence the rotational dynamics, whereas the present work focuses on dilute suspensions.
Their results show that the macroscopic response can be effectively described using an average rod length as a characteristic scale of the system.
In the present study, we demonstrated that for polydisperse systems with a wide and continuous length distribution, using the average value as an ``effective length scale'' allows for the unification of steady-state birefringence and orientation angles measured across various concentrations, which is conceptually consistent with their findings.
Furthermore, we investigate whether this effective length scale can also describe the relaxation dynamics after flow cessation.

\subsection{Relaxation after flow cessation: theory and experiment}\label{sec:RandD_Relaxation}

\begin{figure*}[t]
    \centering
    \includegraphics[width=2\columnwidth]{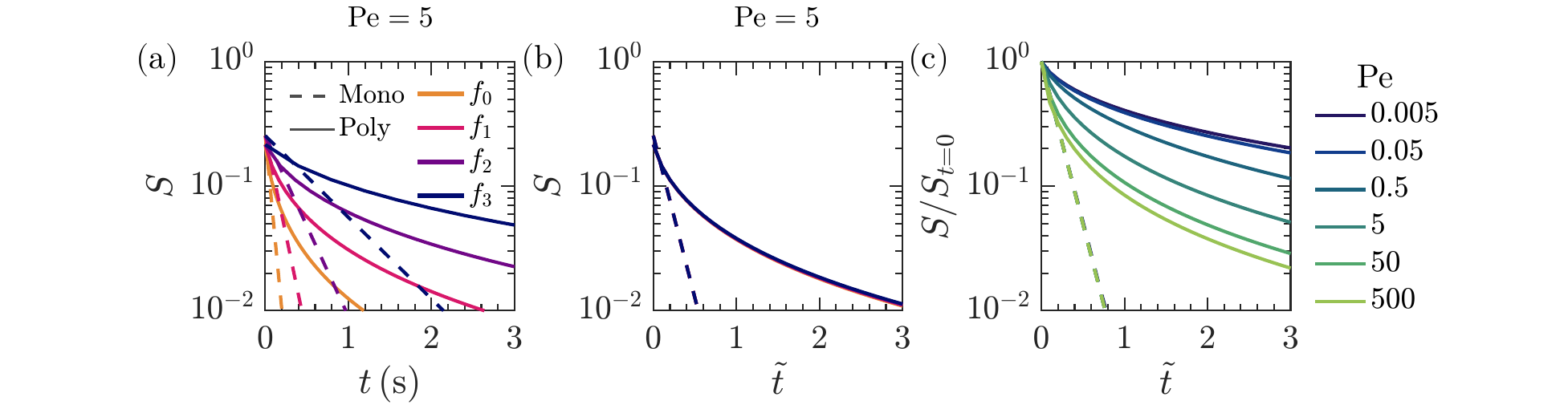}
    \caption{
    (a) \& (b) Decay of order parameter $S$ after shear cessation at $t=0$ against time $t$ and nondimensionalized time $\tilde t = t \Dreff$, respectively, for an initial condition corresponding to $\Pe = 5$ and different weighting parameters $n$.
    (c) Relaxation behavior of the normalized order parameter $S/S|_{t = 0}$ at different initial $\Pe$.
    Dashed and solid lines represent the monodisperse and polydisperse cases, respectively.
    We use an effective rotational diffusion coefficient $\Dreff = D_r(\langle l \rangle_n)$ based on the weighted mean length $\langle l \rangle_n$.
    The polydispersity distributions $f_n(l)$ are based on the fitted number-weighted distribution of CNC rods (see Fig.~\ref{fig:Viscosity_LengthDistribution}(a)), yielding $\exp(\mu) = 238.4\,\mathrm{nm}$ and $\exp(\sigma^2) = 1.3275$.
    The diameter $d = 3.2\,\mathrm{nm}$ is kept constant.}
    \label{fig:S_Relaxation}
\end{figure*}

As pointed out previously, monodisperse suspensions relax exponentially with a single relaxation time $\tau_b = 1/6D_r$ after flow cessation, as illustrated by the dashed curves in Fig.~\ref{fig:S_Relaxation}(a), which show the exponential decay for $S(t)$ with initial $\Pe=5$.
We vary the weighting parameter $n$, following the steady state analysis, and note that larger value of $n$ yield larger $\langle l\rangle_n$ and thus smaller $\Dreff$.
Hence, the decay time in dimensional time $t$ increases with $n$, and the gradient increases (becomes less negative).
By nondimensionalizing the time scale $\tilde t = t \Dreff$, all monodisperse cases collapse onto a single universal exponential decay with expected slope $-6$, shown in Fig.~\ref{fig:S_Relaxation}(b).
The decay is furthermore independent of pre-shear, i.e., independent of the initial $\Pe$, demonstrated in Fig.~\ref{fig:S_Relaxation}(c) by the collapsed dashed lines and consistent with Eq.~\eqref{eq:decay_mono}.
This behavior reflects the fact that orientation relaxes over a finite rotational-diffusion timescale rather than instantaneously.
As a consequence, transient birefringence does not directly follow the instantaneous flow kinematics (for example, the shear rate) in unsteady flows.
This relaxation-induced hysteresis is one reason why direct conversion of transient birefringence to stress via the stress-optic law based on shear rate can be biased in unsteady flows, particularly for long rods or highly viscous solvents.

The polydisperse cases with different weighting parameters $n$ are shown with solid lines in Fig.~\ref{fig:S_Relaxation}(a).
The exponential decay of the monodisperse case is lost and the system exhibits a multi-mode relaxation towards isotropy which is overall slower.
Similar to the monodisperse case, changing $n$ modifies the effective length $\langle l\rangle_n$ and the effective diffusion scale $\Dreff=D_r(\langle l\rangle_n)$ resulting in a slower decay for larger $n$.
Rescaling removes again the strong $n$-dependency and the predicted decays collapse on a single master curve shown in Fig.~\ref{fig:S_Relaxation}(b).
Finally, in contrast to the monodisperse case, the polydisperse decay strongly depends on the applied pre-shear rate, as shown in Fig.~\ref{fig:S_Relaxation}(c) for the normalized $S(t)/S|_{t=0}$.
The decay is slower at low $\Pe$ and becomes progressively faster as $\Pe$ increases, reflecting the fact that the population of rods aligned prior to flow cessation incorporates additional shorter rods with increasing $\Pe$, as already observed in the steady-state analysis, thus decreasing the overall relaxation time for $S(t)/S|_{t=0}$.

\begin{figure*}[t]
    \centering
    \includegraphics[width=2\columnwidth]{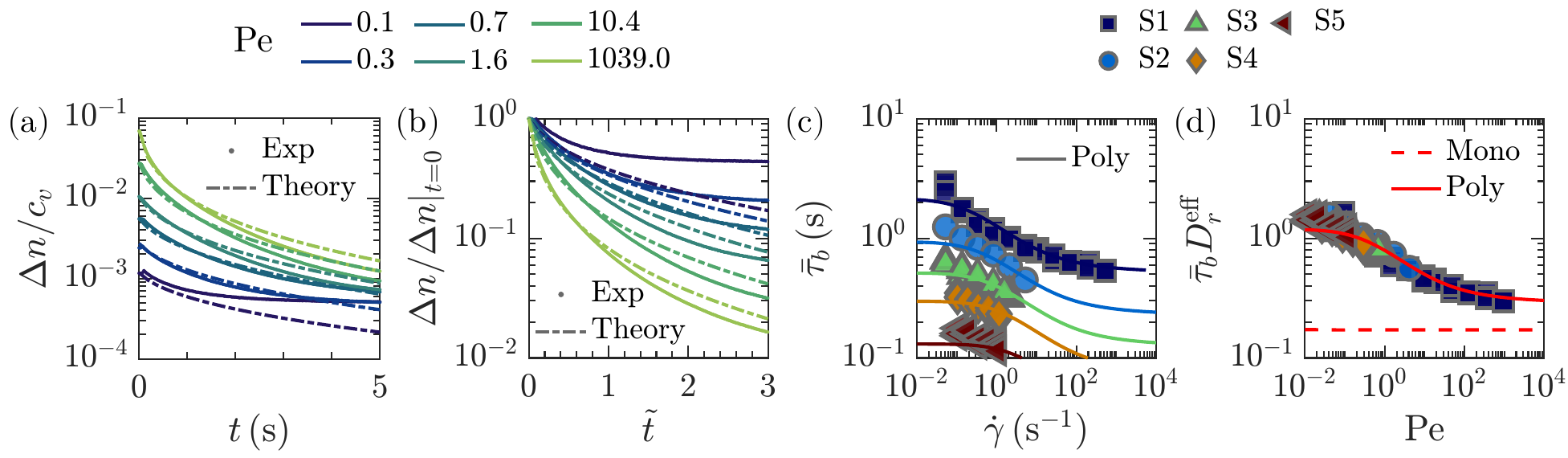}
    \caption{
    (a) \& (b) Decays of concentration-normalized birefringence $\Delta n/c_v$ and birefringence normalized by its initial value $\Delta n/\Delta n|_{t=0}$ after flow cessation at $t=0$, respectively, for $\Pe$ and based on suspension S1.
    Dotted markers denote experimental data and dashed-dotted curves denote theoretical predictions using $n=2$ and $\Delta n_{\rm max}/c_v = 9 \times 10^{-2}$.
    Note that the dotted markers merge into what looks like a solid line due to the high density of data points.
    Each set of experimental decay data represents the average of the data obtained from three experimental trials at the same $\Pe$.
    (c) Orientation relaxation time $\bar{\tau}_b$ as a function of $\dot \gamma$, where $\bar{\tau}_b=\int_{0}^{t_e}\Delta n(t)/\Delta n|_{t=0}\,dt$, evaluated over the observation window up to $t_e$.
    Solid lines represent theoretical predictions for a polydisperse suspension based on the Fokker-Planck model ($n=2$). 
    (d) Dimensionless relaxation time $\bar{\tau}_b \Dreff$ plotted against $\Pe = \dot \gamma/\Dreff$.
    The dashed curve shows the monodisperse prediction, while the solid curve shows the polydisperse prediction.
    }
    \label{fig:relaxation_time}
\end{figure*}

Figure~\ref{fig:relaxation_time}(a) shows the experimentally measured temporal evolution of the concentration-normalized birefringence $\Delta n(t)/c_v$ after flow cessation with different initial $\Pe$, using suspension S1 as a representative example. 
If the suspension were monodisperse, the decay would follow exponential curves independent of $\Pe$ and appear as parallel straight lines on the semi-logarithmic plot as discussed before. 
In contrast, the experimental results deviate systematically: at lower $\Pe$ the birefringence decays slower, whereas at higher $\Pe$ it relaxes more rapidly, reflecting the influence of polydispersity in the rod-length distribution discussed for the theoretical data.
This trend can be seen more clearly by normalizing the birefringence with its initial value and nondimensionalizing the time by the effective rotational diffusion coefficient using $\langle l \rangle_2$, i.e., $\Dreff = D_r(\langle l \rangle_2)$, as shown in Fig.~\ref{fig:relaxation_time}(b).
The dashed-dotted lines in Fig.~\ref{fig:relaxation_time}(a) and (b) represent the theoretical predictions based on the Fokker-Planck model for a polydisperse system using a fixed $n=2$ and $\Delta n_{\rm max}/c_v = 9 \times 10^{-2}$. 
The predictions capture the overall trends of the experimental decays, particularly at early times for higher $\Pe$ cases.
Quantitative deviations remain at later stages of decay, especially for lower $\Pe$.
One possible reason is the weakening of birefringence signals during these stages.
Additionally, for low $\Pe$, the birefringence $\Delta n|_{t=0}$ is already low even before flow cessation.
In particular, $\Delta n$ reaches a minimum measurable birefringence (approximately $1.5 \times10^{-8}$ in our experiment) and does not decrease further (e.g., the early-stage convergence of $\Delta n/\Delta n|_{t=0}$ to a constant value at $\Pe=0.1$ in Fig. \ref{fig:relaxation_time}(b)).

While the full decay curves provide detailed information on the relaxation dynamics, a scalar measure is useful for systematically comparing different flow conditions.
To quantitatively characterize these $\Pe$-dependent relaxation dynamics, we introduce an averaged orientation relaxation time.
Figure~\ref{fig:relaxation_time}(c) summarizes the relaxation dynamics in terms of the averaged orientation relaxation time $\bar{\tau}_b$, obtained by integrating the normalized decay curves in Fig.~\ref{fig:relaxation_time}(b).
As expected by the rotational diffusion coefficient that governs the relaxation behavior of rod orientation, the relaxation time increases with an increase in solvent viscosity.
Furthermore, the relaxation time decreases monotonically with increasing $\dot \gamma$, indicating that samples pre-aligned under stronger shear relax more rapidly once the flow is stopped.
The observed dependence of $\bar{\tau}_b$ on $\dot \gamma$ therefore provides further evidence that the relaxation process reflects the contributions of rods with different lengths. 
In Fig.~\ref{fig:relaxation_time}(d) we nondimensionalize the averaged relaxation time as $\bar{\tau}_b \Dreff$, resulting in a data collapse on a single master curve that matches well the theoretical curve with $n=2$ shown as a solid red line.
The monodisperse prediction (dashed line) corresponds to a constant value of $1/6$ independent of $\Pe$.
We emphasize that the theoretical curves all collapse on the same master curve regardless of $n$, similar to the steady-state flow, but only $n=2$ reproduces the trend with respect to the experimental data (see Fig.~S.3 in the Supplementary Information).
The collapse of the dimensionless relaxation time onto the polydisperse prediction indicates that the observed relaxation dynamics are not governed by a single intrinsic timescale, but emerge from flow-dependent weighting of rod subpopulations with different rotational diffusivities.

Such multi-mode relaxation of birefringence signal in polydisperse rod suspensions has been reported in previous studies \citep{rosenblatt1985,bellini1989,rosen2020,salipante2025}.
In particular, \citet{brouzet2018,brouzet2019} used a flow-focusing channel, in which the local strain rate varies spatially, and reconstructed the rod length distribution from the local birefringence relaxation after flow cessation.
They showed that the relaxation curves depend on the local strain rate immediately before cessation.
The relaxation was faster in regions of high strain rate and slower in regions of low strain rate.
Our observation is qualitatively consistent with this trend, indicating that both polydispersity and flow history play important roles in the orientation relaxation dynamics of rods.

On the other hand, because the flow field in the system of Brouzet et al. was spatially nonuniform, the time during which each rod experienced a given local strain rate was finite.
As a result, especially for longer rods, the flow could shift to another state before sufficient alignment was achieved.
Indeed, their results showed that the reconstructed apparent length distributions were biased toward longer rods in low strain rate regions, whereas in high strain rate regions they were dominated by shorter rods, with little contribution from longer rods.
Therefore, the observed relaxation behavior and the apparent length distributions obtained in their studies were likely influenced not only by the local strain rate immediately before the flow cessation, but also by the prior flow history and the orientational states developed under that history.

In contrast, in our study, the flow field before cessation was simplified to steady simple shear in order to generalize the effect of flow history on the relaxation time.
Under this condition, the suspension can fully develop toward a steady aligned state before cessation, making it possible to define more clearly than in a nonuniform flow which rod populations contribute to the birefringence at each shear rate.
As a result, at low shear rates, the birefringence is governed mainly by longer rods, whereas as the shear rate increases, a broader range of rod lengths, including shorter rods, becomes aligned and contributes to the signal.
This is the essential difference between the studies of Brouzet et al. and our work.
Accordingly, the averaged relaxation time $\bar{\tau}_b$ obtained in this study can be interpreted as a quantity reflecting how pre-shear selects the contributions of different rod populations within a continuous length distribution, and thus can be understood to decrease systematically with increasing $\Pe$.

We should note that our observation of $n=2$ stands in contrast to the weighting parameter $n=3$ proposed by \citet{recktenwald2026} for the same CNC.
However, in agreement with our results (see Fig. S.3 in the Supplementary Information), they find arguably similar agreement between $n=2$ and $n=3$ for steady shear flow.
The difference between the two weighting parameters is most notable in transient flows, with Recktenwald et al. finding generally poor agreement between theory and experiment.
Only under specific flow conditions did both agree, and it is not clear whether $n=3$ consistently outperforms $n=2$ or not.
Moreover, Recktenwald et al. used a histogram for the polydispersity distribution which, due to data sparsity, approximates poorly the long tail of the distribution.
This shortcoming is amplified for higher $n$, thus potentially missing the long tail of the distribution and underestimates rod lengths.
In contrast, studies estimating the length distribution often assume either a length-weighted distribution ($n=1$) \citep{rogers2005, rogers2005a, brouzet2018} or a volume-weighted distribution ($n=1$ or, for electric birefringence, $n=3$) \citep{arenas-guerrero2018, matsumoto1972}.
These studies typically rely on the concept of number concentration \citep{fuller1995}.
However, they also incorporate material properties that depend on weight or length, such as the polarizability anisotropy or the permanent dipole moment in electric birefringence.

Finally, the role of the distribution width, i.e., $\sigma$, is illustrated in Fig.~S.2(d) in the Supplementary Information.
A narrow length distribution yields a relaxation time nearly independent of $\Pe$, whereas broader distributions show a systematic increase in the averaged relaxation time and a pronounced $\Pe$-dependence. 
This behavior arises because long rods dominate the late-time decay when the distribution is broad, leading to slower relaxation compared with the monodisperse limit. 

Overall, Fig.~\ref{fig:BvsPe} and Fig.~\ref{fig:relaxation_time} show that the polydisperse Fokker-Planck model quantitatively reproduces the measured birefringence, orientation angle, and averaged relaxation time when the data are normalized with a single effective diffusion scale $\Dreff$.
Within the tested moment-weighted family, $n=2$ provides the most consistent effective length scale for steady-state alignment and transient relaxation for flow cessation.
These agreements confirm that incorporating a realistic distribution of rod lengths into the Fokker-Planck framework is essential to capture the relaxation behavior of CNC suspensions across different flow conditions.

\section{Conclusion}

In this study, we have shown how polydispersity and flow history jointly control the alignment and relaxation of rod-like cellulose nanocrystals (CNCs) under simple shear.
By combining flow birefringence measurements in a Taylor-Couette geometry with a polydisperse Fokker-Planck model directly parameterized by the measured CNC length distribution, we achieved quantitative agreement between experiments and theory for both steady-state alignment and post-cessation relaxation.

Our key finding is that the averaged relaxation time in polydisperse suspensions is not an intrinsic material constant but depends systematically on the applied pre-shear rate.
While low shear rates primarily align longer rods and lead to slow relaxation, increasing shear rates progressively involve shorter rods, resulting in faster decay.
These results support the hypothesis that in a polydisperse rod suspension, the apparent relaxation time emerges from flow-dependent selection of rod populations with different rotational diffusivities rather than from a single intrinsic timescale.
In this sense, the main conceptual advance of this work is to recast the relaxation time of a polydisperse rod suspension as a flow-history-dependent ensemble property rather than a material-specific constant.

First, steady-state birefringence and orientation angle collapse across concentrations and agree with the polydisperse Fokker-Planck model when plotted against $\Pe=\dot\gamma/\Dreff$ over more than five decades of shear rate.
Here, $\Dreff=D_r(\langle l\rangle_n)$ is an effective diffusion scale defined from a prescribed moment-weighted mean length; among the tested candidates, a single fixed choice $n=2$ provided the most consistent match with theory across concentrations.

Likewise, the averaged relaxation time $\bar{\tau}_b$ decreases systematically with increasing pre-shear rate, and, in the dilute regime, the functional form of $\bar{\tau}_b \Dreff$ as a function of $\Pe$ is independent of concentration but depends on the polydispersity distribution, in contrast to monodisperse systems where $\bar{\tau}_b \Dreff = 1/6$.

Compared with earlier rheo-optical studies that established multi-mode relaxation and inferred rod size information from birefringence transients \citep{bellini1989,brouzet2018}, the present work provides a quantitative link between prescribed steady pre-shear, the measured rod-length distribution and a well-defined averaged relaxation time within a Fokker-Planck framework.
By aligning the suspension to steady-state alignment before cessation, we isolate how pre-shear selects aligned subpopulations and thereby sets the subsequent relaxation dynamics.
These results provide a practical framework for organizing birefringence data in unsteady protocols via a length distribution-informed P\'{e}clet number.

At present, we do not assign a unique physical interpretation to the weighting parameter $n=2$, although prior studies suggest a connection to length- or weight-dependent material properties \citep{rogers2005, rogers2005a, brouzet2018, arenas-guerrero2018, matsumoto1972, reddy2018}.
To physically justify this weighting parameter, systematic experiments using samples with highly controlled size, such as gold nanorods \citep{arenas-guerrero2019,li2014b,perez-juste2005}, would be valuable.

The present framework is restricted to dilute suspensions of rigid rods under simple shear with an effectively constant diameter.
Future work should extend this approach to semi-dilute or concentrated regimes, where rotational diffusion is modified, as well as to more complex flow fields, particularly mixed flows combining extensional and shear components.
Such extensions should further clarify how flow-selected microstructure governs rheo-optical response in broader classes of anisotropic colloidal suspensions.

\section*{Acknowledgments}

This work was supported by JSPS KAKENHI Grant Numbers JP24K07332, JP24K00810, JP24KJ2176 and the Ihara Science Nakano Memorial Foundation FY2024 Research Grant.
VC also acknowledges support from the “la Caixa” Foundation (ID 818 100010434), fellowship code LCF/BQ/PI25/12100026.
The authors also express their sincere gratitude for the financial support provided by the Okinawa Institute of Science and Technology Graduate University (OIST), funded through a subsidy by the Cabinet Office, Government of Japan.

\section*{CRediT authorship contribution statement}

Yuto Yokoyama: Conceptualization, Methodology, Investigation, Data curation, Formal analysis, Funding acquisition, Writing - original draft, review \& editing. 
Vincenzo Calabrese: Conceptualization, Methodology, Funding acquisition, Writing - review \& editing.
Fabian Hillebrand: Methodology, Formal analysis, Writing - original draft, review \& editing. 
Henry J. London: Methodology, Writing - review \& editing. 
Simon J. Haward: Supervision, Resources, Funding acquisition, Writing - review \& editing.
Amy Q. Shen: Supervision, Resources, Funding acquisition, Writing - review \& editing.

\section*{Declaration of generative AI and AI-assisted technologies in the writing process}

During the preparation of this work the authors used generative artificial intelligence tools (ChatGPT, OpenAI) to improve language and clarity.
After using this tool, the authors reviewed and edited the content as needed and take full responsibility for the content of the publication.

\section*{Declaration of competing interest}

The authors declare that they have no known competing financial interests 
or personal relationships that could have appeared to influence the work 
reported in this paper.

\section*{Data availability}

The data that support the findings of this study are available from the 
corresponding author upon reasonable request.

\bibliographystyle{unsrtnat}
\bibliography{ref}

\end{document}